\documentclass[pra,superscriptaddress,twocolumn,notitlepage,showpacs]{revtex4-1}
\usepackage{graphicx} 
\usepackage{amsmath}

\usepackage{lipsum} 
\usepackage[draft]{todonotes}   

\newcommand{\beq}{\begin{equation}}
\newcommand{\enq}{\end{equation}}
\newcommand{\bea}{\begin{eqnarray}}
\newcommand{\ena}{\end{eqnarray}}

\begin{document}
\title{Condensation phenomena in plasmonics}
\author{J.-P. Martikainen, M. O. J. Heikkinen and P. T\"{o}rm\"{a}}
\affiliation{COMP Centre of Excellence, Department of Applied Physics,
Aalto University, P.O. Box 15100, Fi-00076 Aalto, Finland}
\date{\today}

\begin{abstract}
We study arrays  of plasmonic nanoparticles combined with quantum emitters, {\it quantum plasmonic lattices}, as a platform for room temperature studies of quantum many-body physics. We outline a theory to describe surface plasmon
polariton (SPP) distributions when they are coupled to externally pumped molecules. The possibility of tailoring the dispersion in plasmonic lattices allows realization of a variety of distributions, including the Bose-Einstein distribution as in photon condensation~\cite{Klaers2010a}. We show that the presence of losses can relax some of the standard dimensionality restrictions for condensation.  
\end{abstract}
\maketitle

\section{Introduction}
Metallic nanoparticles arranged in periodic arrays
display so-called surface lattice resonances (SLR) with relatively narrow linewidths \cite{Zou2004a,GarciadeAbajo2007,Auguie2008}. SLRs offer complete designability of dispersion using nanofabrication techniques,
which rivals the freedom of tailoring dispersions in optical lattices~\cite{Bloch2008,windpassinger_engineering_2013}, but now in the nanometer scale.
The SLR modes are based on localized surface plasmon resonances (LSPR) of the nanoparticles and diffractive orders of the lattice. The LSPR are confined
to 10-100 nanometer dimensions, providing small mode volumes. Indeed, many-emitter strong coupling has been observed in these and other plasmonics systems~\cite{torma_strong_2014}: 
the normal-mode/Rabi splittings are of the order 100-1000 meV, which means that the light-matter hybrids formed by strong coupling exist 
at room temperature. It has been predicted that strong coupling at the 
{\it single emitter level} is feasible at room temperature and without external cavities~\cite{torma_strong_2014}.
The effective masses of the hybrids in plasmonic systems range typically from  $10^{-8}$ to $10^{-5}$ electron masses, suggesting the possibility of Bose-Einstein condensation at room temperature, for instance (see Appendix~\ref{appendix:effmass}). Thus, we envision that plasmonic lattices combined with emitters, which we call {\it quantum plasmonic lattices}, can become a platform for
quantum many-body physics at room temperature and on-chip. Such similar concepts as superfluid-Mott insulator transition in optical lattices~\cite{jaksch_cold_1998,greiner_quantum_2002}, 
exciton-polariton condensates~\cite{kasprzak_boseeinstein_2006,balili_bose-einstein_2007,plumhof_room-temperature_2013}, light condensation~\cite{Klaers2010a}, thresholdless lasing~\cite{yokoyama_physics_1992,rice_photon_1994}, and quantum fluids of light in general \cite{Carusotto2013} (among others) can be explored in quantum plasmonic lattices. Initial results on lasing in similar systems have been already reported~\cite{oulton_plasmon_2009,zhou_lasing_2013,vanBeijum2013}. 

 The role and design of interactions is  
challenge and an opportunity in quantum plasmonic lattices.
Photon-photon interactions, mediated via either strong coupling hybridization or weak coupling interaction with the quantum emitters, are of a different nature
than in the ultracold gas, semiconductor and microcavity systems mentioned above.
Another difference is the highly lossy character of SLR modes (10-100 fs lifetimes generally), although with SLRs the lifetimes can be potentially increased to 100-1000 fs. A short lifetime is connected to the desirable feature of ultrafast operation speed, but it may hinder interesting quantum and
coherence phenomena. In this article, as a primary example of the physics in quantum plasmonic lattices, we investigate 
the possibility of condensation of surface plasmon polaritons (SPPs) in the SLR modes. We show that by tailoring the SPP losses and 
dispersions 
different SPP distributions including 
one resembling the Bose-Einstein distribution can be realized.
 
\section{Physical model}
\begin{figure} 
\includegraphics[width=0.99\columnwidth]{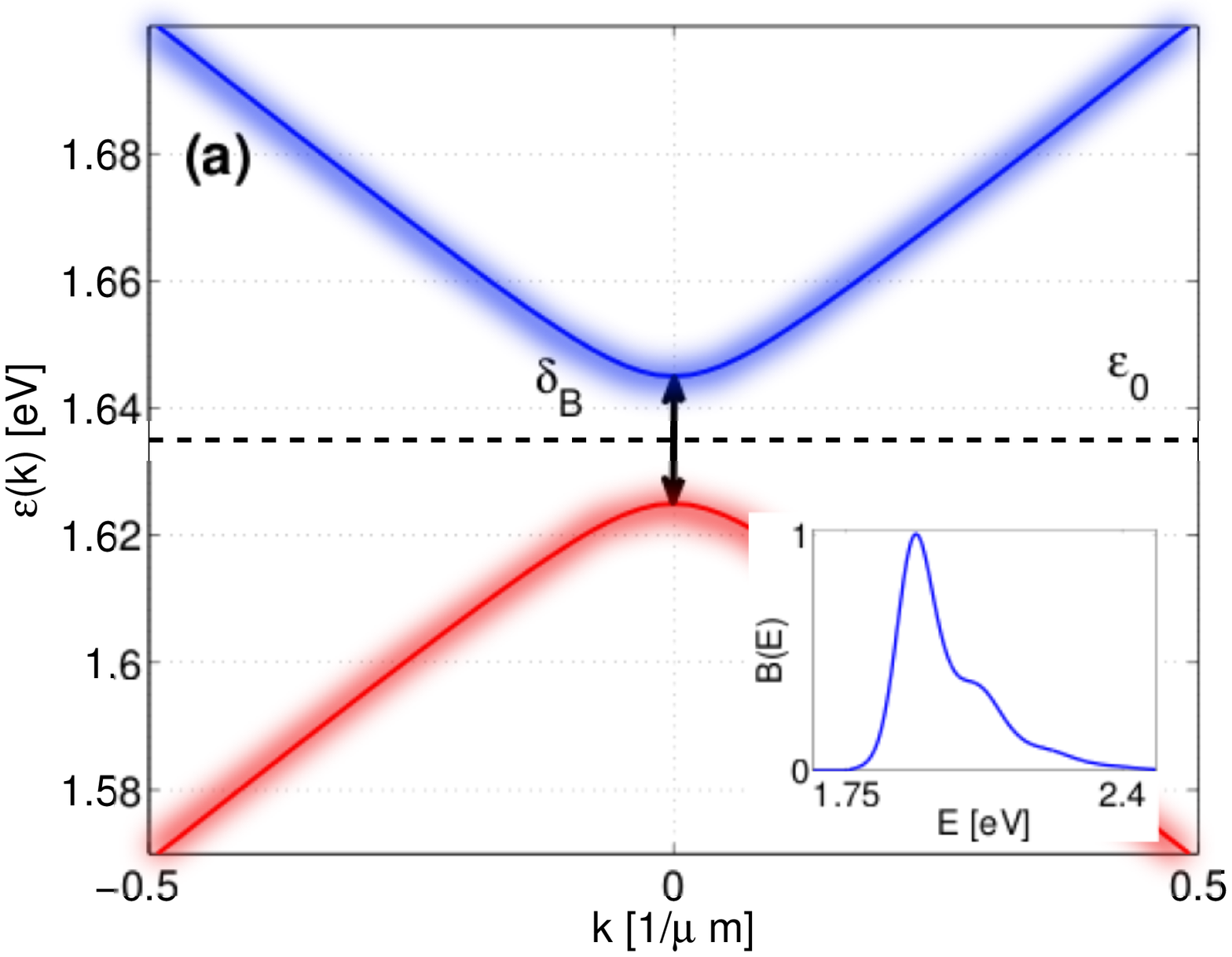}\\
\includegraphics[width=0.9\columnwidth]{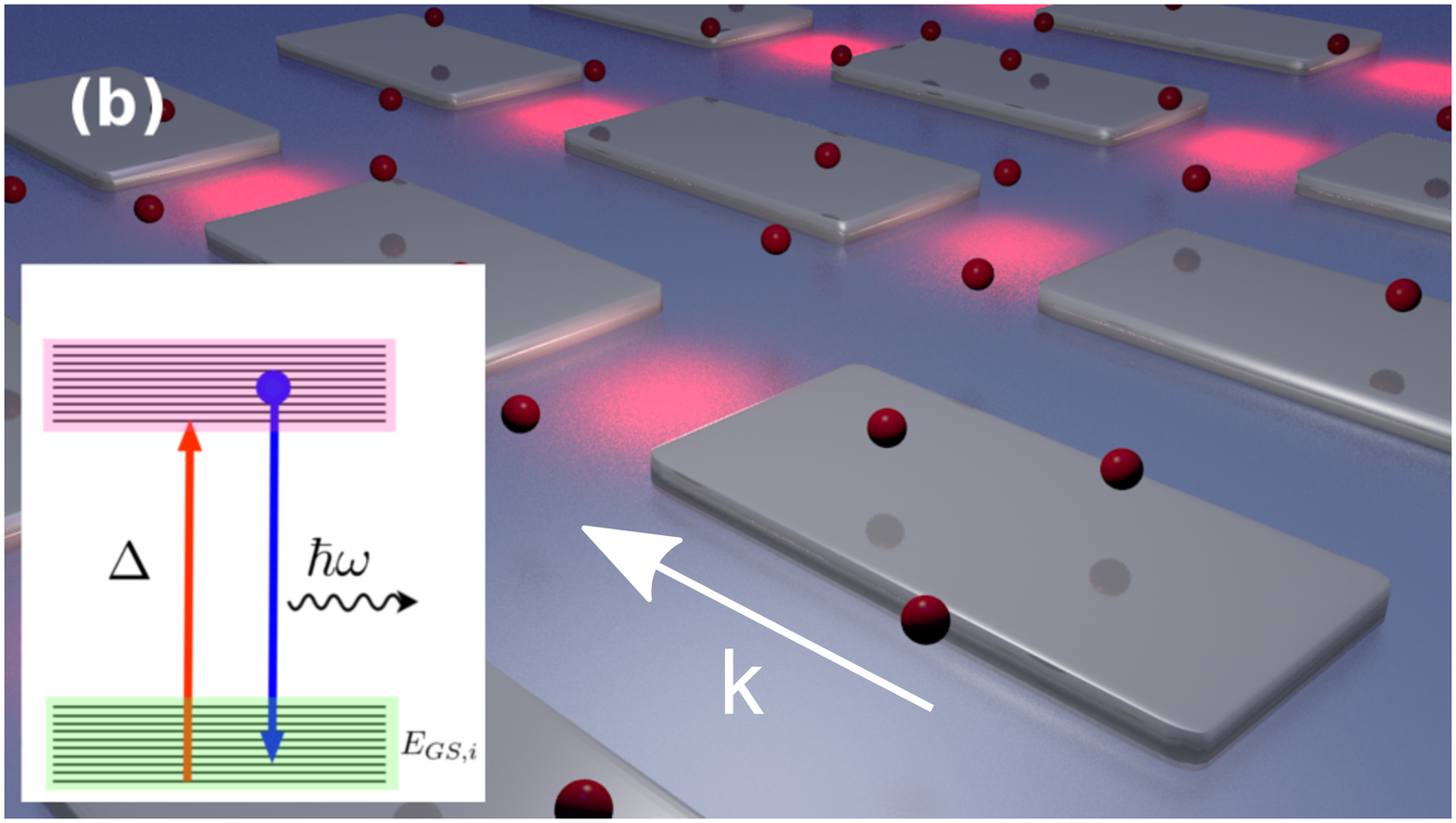}
\caption[Fig1]{SLR dispersions as
a function of wavenumber $k$ in the plane of the nano-particle array. 
Here we choose $\epsilon(k)=\hbar\omega(k)=\pm\sqrt{(\hbar c k/n)^2+(\delta_B/2)^2}+\epsilon_0$, where 
$n=1.51$ is the index of refraction, 
$\delta_B=10\, {\rm meV}$ is a gap between branches, and
$\epsilon_0$ is the energy offset. 
(a) Shows the two SLR branches so that the blurred line on the background indicates finite width of the SLR. In the subplot we
show the absorption profile $B(\omega)$ for the DiD-molecule at
$E=\hbar\omega$
based
on the manufacturer's data together with a linear extrapolation close to
absorption edge.
(b) Shows
a schematic description of the system with rectangular
nano-particles, emitters (red balls), spatially varying electro-magnetic field strength,
and a simplified model of the molecule energy level structure in the inset,
with ground state levels $E_{GS,i}$ ($E_{GS,1}$,$E_{GS,2}$, ...).
White arrow indicates the direction of the wavevector where dispersion varies.
}
\label{fig:schematic}
\end{figure}

We ask can SPPs supported by the SLR modes, interacting with quantum emitters in
the weak coupling regime, display well defined statistical distributions including cases that show condensation ---
even in the presence of high losses? 
We consider a system of metallic nano-particle arrays supporting SLRs, see Fig.~\ref{fig:schematic}. The nanoparticles are anisotropic
so that the SLR dispersions are effectively one-dimensional.
On top of the metallic array structure, we assume a number of quantum emitters that have a
two-level structure where the levels are split into multiple sublevels; in practice
the emitters can be dye molecules embedded in a polymer matrix, for example, with the electronic levels split into a 
rovibrational-level substructure. 
Under appropriate conditions, plasmonic modes can
strongly couple to molecules~\cite{Rodriguez2013,vakevainen_plasmonic_2013,shi_spatial_2014}; here, we consider only
the weak coupling regime. Therefore, we do not consider the possibility of condensation of light-matter hybrids
in analogy to semiconductor exciton-polariton condensates~\cite{kasprzak_boseeinstein_2006,balili_bose-einstein_2007} but phenomena more similar to the concept of photon condensation~\cite{Klaers2010a,fischer_when_2012,kirton_nonequilibrium_2013}.

We consider molecules pumped so that a certain fraction of them are on the excited state manifold. This implies that a loss-compensating 
pump is implicit in our model.
Molecular manifolds thermalize at picosecond timescales due to their
coupling to the phonon bath of the polymer
so that both excited- and ground-state
rovibrational manifolds are separately assumed to be in equilibrium
Maxwell-Boltzmann distributions corresponding to the sample temperature.
Relaxation from the excited to the 
ground-state
is assumed to occur in the timescale of nanoseconds.
We assume the molecular decay is dominated by the coupling to 
the plasmonic modes, that is, a large branching ratio $\beta_b$ between spontaneous emission to the modes of interest versus
other modes~\cite{yamamoto_microcavity_1991,rice_photon_1994}. This is 
justified in plasmonics systems where both enhanced spontaneous emission,
that is, the Purcell effect~\cite{moskovits_surface-enhanced_1985},
and strong coupling can be easily  achieved~\cite{torma_strong_2014,zhou_lasing_2013}. 
Although $\beta_b$ is large, we are not working in the limit of thresholdless lasing~\cite{rice_photon_1994} (see Appendix~\ref{appendix:cavityQED}).
 
In our computations, we consider dye molecules,
although any emitters with a similar level structure would do. 
We take the absorption properties of
the dye molecules to be those of DiD-molecules (as used in our experiments in~\cite{shi_spatial_2014}). 
Importantly (based on the manufacturer data) 
the emission profile is well predicted
by the Kennard-Stepanov relation $\propto B(\omega)e^{-\beta (\hbar\omega-\Delta)}$
at room temperature ($\beta=1/k_BT$). Here $\Delta$ is the difference between the lowest excited
manifold energy and the ground state energy, $\hbar\omega$ is the SPP energy, and $B(\omega)$ is the absorption profile (see Fig.~\ref{fig:schematic}). 
In the limit where
the ground state manifold is narrow, $\Delta$ approaches the absorption edge
of the molecules.
We model the molecules as systems with ground- and excited-state
manifolds that have a rovibrational level structure, see Fig.~\ref{fig:schematic}. We take this level structure 
to be $10$ equidistant energy levels
so that the minimum separation of the manifolds corresponds to the absorption edge.
While realistic molecules are more
complicated, these assumptions do not change the expected qualitative behavior strongly.
The molecules are assumed to be embedded in
a medium, for example, a polymer matrix.

\section{Rate equation model}
Generally time-evolution of a quantum system can be written in terms
of a master equation for the density matrix. 
A simplified description can be obtained by neglecting off-diagonal
coherence terms and focusing on populations appearing on the diagonal.
 These follow rate equations based on transition rates
between states~\cite{mandel_optical_1995,grynberg_introduction_2010}. 

Let us explore the evolution of one $k$-mode (corresponding to SPP
energy of $\hbar\omega(k)$) occupation $n$.
This mode can decay at a rate $\Gamma$, and this gives rise to a loss
term in the rate equation
\beq
{\rm Loss}=-\Gamma n.
\enq
A SPP may also be absorbed by the dye at a rate $R_{abs}$,
and since there are many transitions and the absorption rate depends on the absorption
profile $B$ of the dye,
we get a term
\beq
{\rm Absorption}=-BMR_{abs}n,
\enq
where $M=\sum_i \exp(-\beta E_{GS,i})$ is a factor that accounts for the molecular levels in thermal equilibrium. 
Here we assumed equidistant (splitting 
$\sim 10\, {\rm meV}$) molecular rovibrational levels
over a range of $\sim 100\, {\rm meV}$, but in the end this factor 
depends on the details of molecular-level structure.
Note that in Ref.~\cite{Klaers2011a}, many molecular transitions can contribute to the same photon energy. This was accounted for by an integration over
ground state energies with some rovibrational densities of state. In that case, the steady state could be solved by demanding a vanishing integrand without specifying the densities of states. To find a steady state solution
in our case, the molecules must be
specified in greater detail, since the absorption and emission terms must be 
balanced with the plasmon loss term outside the summation that was absent in Ref.~\cite{Klaers2011a}. Furthermore, we assume that the molecular absorption 
profile depends on the ground state manifold energy $E_{GS,i}$ only weakly
so that $B(\omega,E_{GS,i})\simeq B(\omega)$.

Finally, molecules can decay into SPPs and this process acts as a source
of SPPs. The rate coefficients for spontaneous and stimulated emissions are the same.
This similarity together with the assumption of the Kennard-Stepanov law relating
emission and absorption rates and the $M$ factor mentioned earlier, gives a term
\beq
{\rm Emission}=MBR_{spon}\left(1+n\right)e^{-\beta (\hbar\omega(k)-\Delta)}.
\enq
The ratio of $R_{spon}$ and $R_{abs}$
depend on the number of excited and ground state molecules, as will be discussed below.

As a result of these considerations
SPP occupations $n(k)$ are described by a rate equation 
\begin{eqnarray}
{\dot n}(k)&=&-\Gamma(\omega(k)) n(k)+\sum_i B(\omega(k))e^{-\beta E_{GS,i}}\times \label{fullrateq} \\
&&\left[R_{spon}(\omega(k))\left(1+n(k)\right)e^{-\beta (\hbar\omega(k)-\Delta)}
\right.\nonumber\\
&-&\left.R_{abs}(\omega(k))n(k)\right]\nonumber.
\label{eq:rateq}
\end{eqnarray}


Here, $\omega(k)$ is the SLR dispersion. The summation is over ground state molecular manifold energy levels $E_{GS,i}$.
We estimate the SPP decay rate $\Gamma(\omega)$ from our experiments~\cite{shi_spatial_2014}. With reasonable accuracy
it can be taken as a constant of magnitude $\Gamma(\omega)=10\, {\rm meV}$.
This term, addressing the
short lifetimes of SPPs, is one of the essential differences 
when compared to previous studies
of photon condensation in microcavity systems~\cite{klaers_boseeinstein_2011}
where such a term was ignored due to 
longer lifetimes.
The remaining terms
describe spontaneous and stimulated 
emissions at rate $R_{spon}(\omega)$ 
from molecules to SPPs as well as an absorption 
at the rate $R_{abs}(\omega)$.
Note that chemical potential does not appear in the rate equation since the SPP number is not conserved.

The number of molecules $N_M$ coupled to the
SPPs should not be obtained 
from the total number of molecules in the sample, but instead 
as the number of molecules within an effective mode volume; this is
estimated (see Appendix~\ref{appendix:molnum}) to be 
around $N_M=24\cdot 10^6$ dye molecules in our system. 
We define $N_M = N_e + N_g$,
where $N_e$ is the number of molecules in the excited manifold and $N_g$ corresponds
to the molecules in the ground state.

The rates are affected by the number of excited state
molecules $N_e$ relative to ground state molecules $N_g$, and the quantum
yield $\Phi$ of the dye-molecule.  
We take the rate coefficient for the spontaneous (and
stimulated) emissions to be
\beq
R_{spon}(\omega)=\frac{F}{\tau} N_{e},
\label{eq:rate}
\enq
where $\tau$ is the characteristic lifetime of the excited state molecule
in vacuum, chosen as $\sim 5\, ns$. In principle, the spontaneous emission rate in free space depends on energy as $\propto \omega^d$ ($d$ is the dimensionality), 
but we ignore the energy dependence
here since the range of energies where distributions vary is small
compared to absolute photon energies and the presence of metal will change such dependencies. Furthermore, $F$ is the Purcell factor
and describes the enhanced decay rate of the molecules close to
plasmonic structures. 
Interestingly, large Purcell enhancement will 
allow smaller population inversions, that is, smaller $N_e$.
Since $F$ is unknown and depends on experimental details, we choose $F=20$ in our computations so that it is likely of correct order of magnitude with respect to the range of reported values from $F=6$ to $F=200$~\cite{oulton_plasmon_2009,zhou_lasing_2013}.
The absorption rate coefficient is taken as 
\beq
R_{abs}(\omega)=N_g R_{spon}(\omega)/(N_e\Phi).
\enq 
Note that since the total absorption rate scales with the number of molecules the rate is, in our case, typically few orders of magnitude larger than the SPP decay rate.

The SPPs with energies in the $2\, {\rm eV}$ range
are well separated from thermal excitations
at $k_BT=26\, {\rm meV}$, so the only way to excite them is via decay from molecules.
The SPPs can have fairly 
complicated dispersions, and their widths can vary, for example,
as a function of the photon momenta in the plane of the nano-particles.
The dispersions are typically nearly linear, however a bandgap can be designed
by breaking the symmetry of the array, see Fig.~\ref{fig:schematic}. 
It should be emphasized that SLR dispersions can 
be tailored at will with different structures,
and our choices are therefore only indicative of the physically-possible dispersions. 
The anisotropy of the array implies that SPPs of 
only one polarization state play a role here, and the dispersion is essentially one-dimensional~\cite{rodriguez_coupling_2011}.
The SLR dispersion used here is, as given in the caption of Fig.~\ref{fig:schematic},
\begin{equation}
\epsilon(k)=
\pm\sqrt{(\hbar c k/n)^2+(\delta_B/2)^2}+\epsilon_0.
\end{equation}
Here $k$ is the wavenumber in the plane of the nano-particle array, 
$n$ is the index of refraction, 
$\delta_B$ is the gap between the upper and lower 
SLR branches, and
$\epsilon_0$ is located in the middle of the dispersions. We use the
typical values $n=1.51$ and 
$\delta_B=10\, {\rm meV}$, and $\epsilon_0$ is varied.

\section{Steady state solution}
The rate equations have a steady state solution 
\beq
n(k)=\frac{1}{g(\omega)e^{\beta(\hbar\omega-\Delta)}-1},
\label{eq:steadystate}
\enq
where we defined 
\beq
g(\omega(k))=[MB(\omega)R_{abs}+\Gamma(\omega)]/(M R_{spon}B(\omega))
\enq 
and $M=\sum_i \exp(-\beta E_{GS,i})$. 
Now $k$-states
for which the denominator (nearly) vanishes may have, for certain parameters, diverging populations compared to the other $k$ states
i.e.\ macroscopic occupation.
If the term $g(\omega)e^{\beta(\hbar\omega-\Delta)}$ is much larger than unity, then a distribution similar to 
the Maxwell-Boltzmann one is obtained. To have
diverging populations of the lowest states,
one should have $g(\omega)e^{\beta(\hbar\omega-\Delta)}$ around $1$ for small energies. 
This implies $R_{abs}/R_{spon}=N_g/(N_e\Phi)\sim 1$. Note that if molecules experience losses to other modes than SLRs, these could be taken into account by reducing the quantum yield $\Phi$.

\begin{figure} 
\includegraphics[width=0.95\columnwidth]{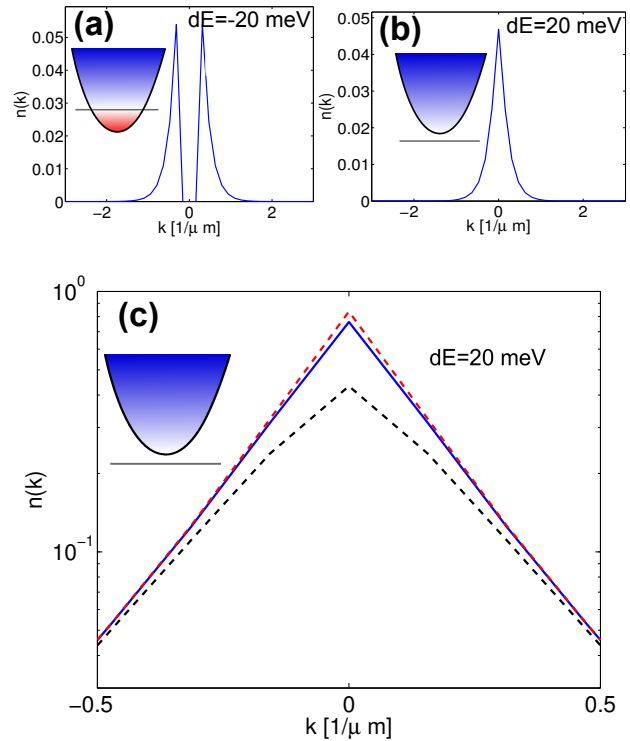}
\caption[Fig2]{SPP occupation numbers in 
the upper branch at
room temperature as a function of in-plane momentum when losses were $\Gamma(\omega)=10\, {\rm meV}$. 
Depending on the relative positions of the molecule absorption edge and the dispersion band edge, one observes
accumulation of population to different $k$-states. 
In (a) and (b) we choose $N_e/N_g=0.1$ and different figures correspond to different 
separations of the absorption edge of the molecules from the minimum 
of the SLR dispersion $dE=min[\hbar\omega(k)]-\Delta$. The hole in first
distribution is caused by
absorption edge of the molecules
being at higher energy than the minimum of the SLR dispersion. Position
of the absorption edge is indicated schematically in the insets 
by the gray line. Only
modes in the blue region are appreciably pumped by molecules. As temperature increases this feature becomes less sharp.
 In (c) we approached
inversion so that $N_e/N_g=0.99$. Solid line is from our theory,
dashed black line shows the Maxwell-Boltzmann 
result without "-1" in the denominator of the distribution, and
dashed red line is with a Bose-Einstein distribution where $g(\omega)$ 
was evaluated close to absorption peak at $1.92\, {\rm eV}$.
}
\label{fig:1Ddistribution}
\end{figure}

To have the term minus one in the denominator of equation (\ref{eq:steadystate}),
the term $n(k)$ in $(1+n(k))$ in the rate equation (\ref{fullrateq}) is needed: this means having stimulated
emission into the mode of interest. Bosonic enhancement is thus behind any SPP condensation with statistics deviating from
the Maxwell-Boltzmann distribution.
The spontaneous emission term, the unity in
$(1+n(k))$, is needed to have a non-zero SPP number, since there is no direct pumping into the mode (see Appendix~\ref{appendix:cavityQED}). Losses play a role somewhat similar to absorption. In the limit of small
losses ($\Gamma$) and absorption ($R_{abs}$), the distribution (\ref{eq:rateq}) becomes negative, meaning there is no steady state and only
exponentially growing solutions;
we differentiate condensation from lasing 
by exponentially growing $n$, that is, a positive gain coefficient for the latter. 
In Fig.~\ref{fig:1Ddistribution}, we demonstrate the distributions for realistic experimental parameters.

\begin{figure}
\includegraphics[width=0.8\columnwidth]{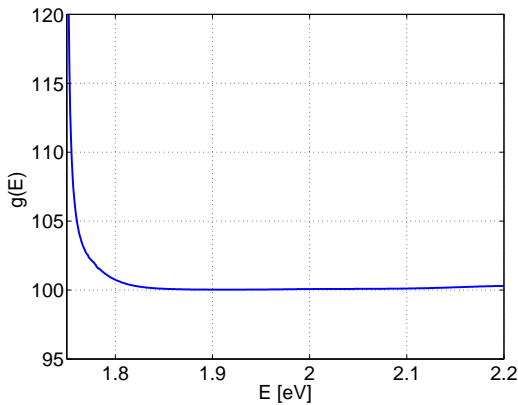}
\caption[Fig2]{The function $g(\omega)$ in our example
system with $N_e/N_g=0.01$ ($E=\hbar\omega$). 
For the DiD-dye molecules the absorption edge is at $\Delta=1.75\, {\rm eV}$
and we choose $\Gamma(\omega)=10\,{\rm meV}$ which is a reasonable
estimate based on experiments~\cite{shi_spatial_2014}.
For a BE-distribution this function should be constant.}
\label{fig:gfunc}
\end{figure}

How close to a Bose-Einstein distribution can one make 
the distribution? By moving the function $g(\omega)$ 
in Eq.~(\ref{eq:rateq}) into the exponent
we find a distribution with the "chemical potential" $\mu(\omega)=\Delta-\frac{1}{\beta}\ln g(\omega)$ that generally depends
on energy. 
In the special case where losses disappear $\Gamma(\omega)=0$ the result
becomes similar to that for the chemical potential of photons in a 
photon BEC~\cite{Klaers2011a} where photon fugacity was essentially fixed by the excitation level of the system.
More generally it should be noted that 
when $\Gamma(\omega)/B(\omega)$ 
is constant (other quantities in Eq.~\ref{fig:gfunc} are assumed constant) $\mu(\omega)$ is also a constant  and the distribution still looks like a Bose-Einstein distribution even
in a lossy environment.
This is interesting considering the freedom of tailoring the dispersion in plasmonic lattices.
Like dispersions loss profiles can also be designed by the lattice geometry. Therefore, it is possible to 
engineer loss profiles to match the absorption profile of the emitters, and thus perhaps guarantee
a constant  $\Gamma(\omega)/B(\omega)$. One possibility to do this is to couple the
SLR of interest to another resonance (e.g. caused by another diffractive order in the lattice)
so weakly that no splitting but merely a broadening of the dispersion is caused near the desired
location in energy. 
 
\begin{figure}
\includegraphics[width=0.8\columnwidth]{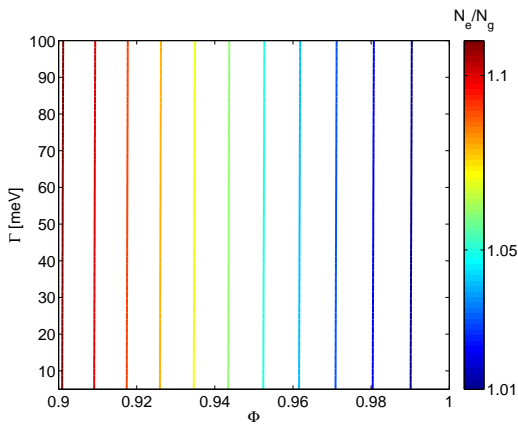}
\caption[Fig1]{The value of $N_e/N_g$ required for having diverging
SPP occupation number. We kept other parameters the same as elsewhere,
but allowed for changes in the quantum yield $\Phi$ of the emitter and
the loss rate $\Gamma$. (Energy was fixed at $1.92\, eV$ where the energy
dependence of the absorption profile is weak.)}
\label{fig:BElimit}
\end{figure}

Furthermore, Bose-Einstein-like distribution might appear over some energy intervals but not in others.
In Fig.~\ref{fig:gfunc} we
show an example of how the function $g(\omega)$ behaves when the loss coefficient
is constant ($\Gamma(\omega)=10\,{\rm meV}$)
and the absorption profile corresponds to DiD-molecule. As can be seen this function is almost constant over
a broad range and has strong energy dependence only close to the absorption
edge where the denominator of $g(\omega)$ approaches zero.
Beyond this regime, 
around the absorption peak, $g(\omega)$ is almost constant for an
energy range much broader than room temperature; it is
in this range where Bose-Einstein distribution may be expected 
even without special tailoring of the loss profile, as shown in Fig.~\ref{fig:1Ddistribution}. The condition for diverging populations is 
\beq
N_e\Phi/N_g\ge \left(1-\Gamma(\omega)\tau/(MFN_eB(\omega))\right)^{-1}.
\enq 
As mentioned earlier, in case of a constant $\Gamma(\omega)/B(\omega)$ the resulting distribution
is the BE one, otherwise it deviates from BE-distribution, but there
is nevertheless divergence of population.
We demonstrate the way excitation fraction required for
diverging occupation behaves as a function of quantum yield and
loss rate in Fig.~\ref{fig:BElimit}. With our parameters this quantity
depends only relatively weakly on the SPP loss coefficient while the dependence on quantum yield is stronger.

The theory of ideal Bose-Einstein condensates (BEC)
with dispersions linear in $k$ shows that a BEC cannot occur
in a one-dimensional system. Likewise it is not possible with free particle
like $k^2$-dispersions in dimensions less or equal to two~\cite{pethick_bose-einstein_2008}.
In our context, such limitations can be relaxed since the
loss profile and the absorption profile 
both appear in the steady state solution.
To illustrate this, assume $\epsilon=\hbar \omega = \hbar c k/n$. We can then write the total steady-state SPP number ($L$ is the length of the system)
\beq
N_{SPP}=\sum_k n(k)=\frac{L}{2\pi}\int dk\, n(k)=\frac{nL}{hc}\int \frac{ d\epsilon}{g(\epsilon)e^{\beta(\epsilon-\Delta)}-1}\nonumber
\enq
also in terms of effective density of states as
\beq
N_{SPP}=\int d\epsilon \frac{\rho(\epsilon)}{e^{\beta(\epsilon-\Delta)}-1},
\enq
where
\beq
\rho(\epsilon)=\frac{nL}{h c}\frac{e^{\beta(\epsilon-\Delta)}-1}{g(\epsilon)e^{\beta(\epsilon-\Delta)}-1}.
\enq
This function is not the physical density of states,
and consequently BEC-like distributions can appear 
even in lower dimensional systems if the
loss behavior is appropriate. 
In this case close to the absorption edge,
$\rho(\epsilon)$ vanishes like the density of states for a three-dimensional photon gas or for massive 
particles in a three-dimensional harmonic trap (see Appendix~\ref{appendix:densityofstates}). The integral over energy is therefore
well behaved in a similar way 
as in the theory of the ideal BEC.

\section{Lasing}
Finally, we connect the results to lasing in these systems. For exponentially 
growing solutions of equation (\ref{fullrateq}) to occur the gain coefficient
\beq
\alpha(\omega)=\frac{MB(\omega)\left[R_{spon}(\omega)e^{-\beta(\hbar\omega-\Delta)}-R_{abs}(\omega)\right]}{\Gamma(\omega)}-1
\nonumber
\enq
should be positive.
This is only possible if
\beq
\frac{N_e\Phi}{N_g}> e^{\beta\hbar (\omega-\Delta)}
\enq
which implies population inversion.
This is not a sufficient condition for
positive gain. While close to the absorption edge, the exponential
approaches one; the $B(\omega)$ multiplier in the gain coefficient approaches
zero.

\begin{figure}  
\includegraphics[width=0.95\columnwidth]{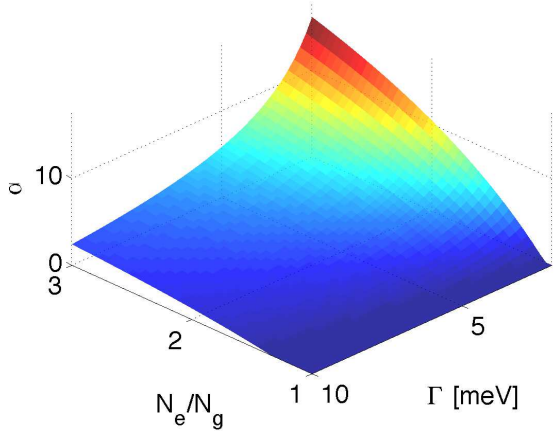}
\caption[Fig3]{The maximum of the gain coefficient $\alpha(\omega)$ as a function of inversion $N_e/N_g$ and the loss rate. We choose
$\Delta=1.75\, {\rm eV}$, 
temperature $T=300\, {\rm K}$, and
a Purcell factor of $F=20$.
}
\label{fig:gaindemo}
\end{figure}

In Fig.~\ref{fig:gaindemo}, we demonstrate the behavior of the maximum of
the gain coefficient. 
The parameters of the experiment~\cite{zhou_lasing_2013} that reported an observation of lasing 
in plasmonic arrays correspond roughly to
the left corner of Fig.~\ref{fig:gaindemo}.  In our example,
the maximum gain there is indeed positive 
($\alpha\approx 2.4$), which means that our model 
is consistent with existing experiments.

\section{Conclusions}
We have shown that condensation phenomena are possible even in 
high loss rate systems, such as plasmonic lattices combined with
emitters. Population inversions of $N_g/(N_e\Phi)\sim 1$ are needed to achieve the regime where the distribution
deviates from the Maxwell-Boltzmann one and may show diverging occupation numbers for certain states. 
By tailoring the dispersions and loss profiles, different types of distributions can be produced, also the Bose-Einstein one. 
Coherence and quantum statistical properties of these systems should be characterized in future work, including higher-order correlations beyond the rate equation 
approach~\cite{schmitt_observation_2014,altman_two-dimensional_2013}. 
Condensation of SPPs may be the way to produce highly coherent light in nanoscale
with low powers and modest population inversions.

\begin{acknowledgments}
We thank Jildou Baarsma, Tommi Hakala, Robert Moerland, Heikki Rekola, and Lei Shi for useful discussions.
This work was supported by the Academy of Finland through its
Centres of Excellence Program (Projects No. 251748, No. 263347 and No. 13272490) and by the
European Research Council (ERC-2013-AdG-340748-CODE)
\end{acknowledgments}

\appendix
\section{Estimates for the effective mass}
\label{appendix:effmass}
The interplay between diffraction orders, localized surface plasmon resonances, and
emitters 
may introduce splittings of the surface lattice resonances. 
In the absence of coupling to emitters, surface lattice resonances can have (for example) a dispersion of type
\beq
\epsilon(k)=\epsilon_0\pm\sqrt{(\hbar c k/n)^2+(\delta_B/2)^2}, 
\enq
where $n$ is the index of refraction, 
$\delta_B$ is a gap between the branches, and
$\epsilon_0$ is the energy offset. For small wave-vectors $k$ in the 
plane of the nano-particles, we can 
use a Taylor expansion of 
the square root and find a particle-like dispersion that is described by an effective mass
\beq
m_{eff}=\frac{n^2\delta_B}{2c^2}.
\enq
When the splitting is in the range from
$10\, {\rm meV}$ to $1000\, {\rm meV}$,
the effective
mass is between $10^{-8}\, m_e$ 
and $10^{-6}\, m_e$,
where $m_e$ is the electron mass. Since, for example, the critical temperature
for Bose-Einstein condensation is inversely proportional to particle mass
($k_bT_c\approx 3.3 \hbar^2 n_B^{2/3}/m_{eff}$ for a free Bose gas in 3D with density $n_B$), these
low values suggest a possibility for much higher temperatures for phase transitions.
However the above approximation is not valid for all wave-vectors, and
real dispersions should be used in actual calculations. Expansion is appropriate
if $k\ll n\delta_B/(2\hbar c)$ where the right-hand side varies ($n=1.5$) between
$0.04/{\rm \mu m}$ and  $4/{\rm \mu m}$.

We can also estimate the effective mass where the above
dispersions couple strongly with a molecular energy level of 
the emitter~\cite{vakevainen_plasmonic_2013,shi_spatial_2014}. In this case, the eigenmodes follow from a Hamiltonian
\beq
H=\left(\begin{array}{cc}
\sqrt{(\hbar c k/n)^2+(\delta_B/2)^2} & \Omega\\
\Omega^* & E_m
\end{array}
\right),
\enq
where $E_m$ corresponds to the molecular transition energy 
and $\Omega$ is the coupling between
molecules and the surface lattice resonance (without a loss of generality, we choose $\epsilon_0=0$ in this case).
Solving for the eigenmodes and expanding again around $k=0$, gives 
us the effective mass
\beq
m_{eff}=\frac{n^2\delta_B \sqrt{(2E_m-\delta_B)^2+16|\Omega|^2}}
{c^2\left[2E_m-\delta_B+\sqrt{(2E_m-\delta_B)^2+16|\Omega|^2}\right]}.
\enq
This approaches the previous result as coupling disappears, while in the limit
of very large coupling, we have $m_{eff}\approx n^2\delta_B/c^2$. The effective mass of the hybridized mode is nevertheless always larger than 
that of the surface lattice resonance. If 
the splittings in SLR range from $\delta_B=10\, {\rm meV}$ 
to $\delta_B=1000\, {\rm meV}$ 
and the Rabi splitting is $100\, {\rm meV}$~\cite{shi_spatial_2014},
the effective
mass varies between $m_{eff}\sim 10^{-8}\, m_e$
and $m_{eff}\sim 10^{-5}\, m_e$.

\section{Comparison to cavity-QED laser rate equations}
\label{appendix:cavityQED}
It is instructive to compare the rate equation~(\ref{eq:rateq}) to the description of
a cavity-QED laser~\cite{rice_photon_1994}. The rate equation of the cavity-QED
system in the presence of background absorption is
\begin{align}
\gamma^{-1}\dot{n}&=-\lambda n+\beta_b n (N-N_0)+\beta N,\nonumber\\
\gamma^{-1}\dot{N}&=-N+P-\beta_b n (N-N_0).
\end{align}
Here, $n$ is the photon number in the cavity mode and $N$ is the number of atoms in the
upper lever of the lasing transition, in short, the number of carriers.
The coefficients $\lambda$ and $P$ are the cavity decay rate and the pumping rate, respectively,
measured in the units of the spontaneous emission rate $\gamma$.
The branching ratio $\beta_b$ indicates the fraction of the spontaneous emission
to the laser mode~\cite{yamamoto_microcavity_1991}. Finally, $N_0$ is the carrier number at transparency and accounts for
the absorption of photons to the medium
(more precisely, when $N=N_0$, the stimulated emission from and absorption to the medium are equal).

The pair of equations above assumes that the system is pumped with a constant rate $P$.
This assumption is the important difference in the rate equation~(\ref{eq:rateq}) where,
conversely, it was assumed that the excited molecular states are in thermal equilibrium with
the environment, and that the time-scale of equilibration is fast in comparison to the time-scale
of the rate equation. This is equivalent to replacing the time-evolution of the carrier density
of the cavity-QED system with an externally-determined, constant carrier density $N_c$.
Incorporating this assumption, only the
photon number rate equation is relevant to the problem. We re-write this equation as
\begin{align}
\dot{n}&=-\gamma\lambda n+\gamma\beta_b N_c (1+n) -\gamma\beta_b N_0 n.
\end{align}
We may then readily identify a one-to-one correspondence with the SPP rate equation.
The coefficients of the two rate equations map onto each other as follows
\begin{align}
\gamma\lambda &\leftrightarrow \Gamma(\omega(k)), \\
\gamma\beta_b N_c &\leftrightarrow 
\left(\sum_i B(\omega(k)) e^{-\beta E_{GS,i}}\right) R_\textrm{spon}(\omega(k)) e^{-\beta[\hbar\omega(k)-\Delta]},\\
\gamma\beta_b N_0 &\leftrightarrow \left(\sum_i B(\omega(k)) e^{-\beta E_{GS,i}}\right) R_\textrm{abs}(\omega(k)).
\end{align}
In other words, the cavity decay rate corresponds to the SPP decay rate,
the total emission rate to the cavity corresponds to the total emission rate to the SLR mode,
and the background absorption rate corresponds to the absorption of SPP by the molecules.
The difference is that the rate equation for the SPP system explicitly 
takes into account that
there are many molecular energy levels which take part in the emission and absorption processes,
that is, there is a summation over the molecular energy levels $E_{GS,i}$.

There are number of important points to note. Several SLR modes are labeled by the
momentum $k$, each of which requires its own rate equation. 
However, these rate equations are mutually independent since the molecular states are maintained
at thermal equilibrium with constant occupation number.
Therefore, the mapping to the cavity-QED system is possible on the level of a single SLR mode. 
The high loss rate character of our system guarantees a small SPP number
compared to the number of excited molecules; therefore, the assumption
of the constant occupation number is reasonable.
Thus, although the $\beta_b$-factor in our case is large, that is, 
most of the spontaneous emission is guided to the modes of interest, we are not working in 
the regime of thresholdless lasing. Our system rather corresponds to the bad cavity limit. 
Indeed, if the molecular occupation numbers would be time-dependent and have feedback from the SPP spectrum,
all of the SLR modes would be coupled with each other 
by the time evolution of the molecular states. In summary,
the difference of our case in comparison to cavity-QED system 
description~\cite{rice_photon_1994} are 1) constant excited state 
(carrier density) population assumed instead of excited state dynamics,
which is justifiable in our high-loss-rate system, 2) we consider several
modes, 3) temperature- and energy-dependent excited state
populations and absorption/emission coefficients. Due to these differences,
phenomena distinct from (thresholdless) lasing can be predicted.

\section{Estimate of the molecule number}
\label{appendix:molnum}
The rates in the rate equations 
are affected -- among other things-- by the total number of molecules $N_T$. The number of molecules $N_M$ coupled to the
SPPs will, however, be different from the actual number of molecules
over the sample for several reasons. First, the mode structure of the SLR
modes has a strong spatial variation at a length scale of the metallic
nanoparticle array. This means that the molecules around "hot spots" 
dominate the coupling to SPPs. The fraction of such molecules
is roughly proportional to the fraction of the surface covered by metal,
and in our typical experiments~\cite{shi_spatial_2014} this varies between
$0.3\%$ and $1.3\%$. Second, the intrinsic losses in the system limit how
far the modes can propagate, and coherence lengths related to this are in the
range of $\sim 5\, {\rm \mu m}$~\cite{shi_spatial_2014}, which is much smaller
than the dimensions of the nanoparticle array. Together, this suggests that 
$N_M$ should range from $0.1 \%$ to $1 \%$ of the total number of
dye-molecules. We choose an estimate of $1\%$. 

We consider a molecule concentration of $50\, mM$, and $50\, {\rm nm}$
as the thickness of the molecule layer. For a $40\times 40\, {\rm \mu m}^2$
sample, this implies
around $N_M=24\cdot 10^6$ dye molecules. This is a reasonable value
based on our experiments~\cite{shi_spatial_2014}
where the dye concentration range from
$0$ to $800\,{\rm mM}$, and up to $1\,\%$ of 
the structure was covered by metal.

\section{Effective density of states}
\label{appendix:densityofstates}
\begin{figure}
\includegraphics[width=0.8\columnwidth]{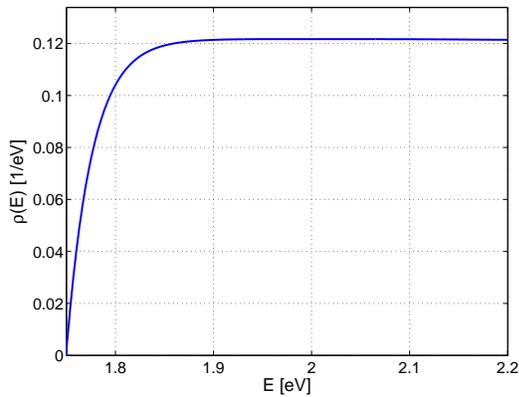}
\caption[Fig2]{Function $\rho(E)$ with linear SLR dispersions which cross
at the absorption edge, $\Delta$, of the molecules. We used $\Delta=1.75\, {\rm eV}$, $T=300 K$, and refractive index $n=1.51$.
(For concreteness we choose the length scale in the prefactor in Eq.~(\ref{eq:rho}) to be $L=10\,\mu m$.)}
\label{fig:rhofunc}
\end{figure}

As pointed out in the main text, with linear dispersion we can write 
the plasmon number in terms of effective density of
states as
\beq
N_{pl}=\int d\epsilon \frac{\rho(\epsilon)}{e^{\beta(\epsilon-\Delta)}-1},
\enq
where
\beq
\rho(\epsilon)=\frac{nL}{h c}\frac{e^{\beta(\epsilon-\Delta)}-1}{g(\epsilon)e^{\beta(\epsilon-\Delta)}-1}.
\label{eq:rho}
\enq
Close to the absorption edge we can take $B(\epsilon)\propto (\epsilon-\Delta)$. Since
the loss coefficient $\Gamma(\omega)$ is non-zero and expected to be 
well behaved at the absorption edge,
this implies 
$g(\epsilon)=a+b/(\epsilon-\Delta)$ (with $a$ and $b$ some positive coefficients) in the vicinity of the absorption edge. Therefore, 
by expanding $\rho(\epsilon)$ around the absorption edge we find
$\rho(\epsilon)=nL\beta (\epsilon-\Delta)^2/(bhc)$. As a function of energy relative to absorption edge,
this function  vanishes in the same way as the density of states for a three-dimensional photon 
gas or similar to the density of states of massive 
particles in a three-dimensional harmonic trap. 
We demonstrate this behavior in Fig.~\ref{fig:rhofunc}.
If the absorption profile close to the absorption edge vanishes faster than linearly,
then the effective density of states vanishes faster than $(\epsilon-\Delta)^2$ and the integration over energy is still well behaved. Note that in the figure the region where $\rho(\epsilon)\propto (\epsilon-\Delta)^2$ is so close to the absorption edge that the 
dependence appears roughly linear visually.

\bibliographystyle{apsrev}

\begin{thebibliography}{33}
\expandafter\ifx\csname natexlab\endcsname\relax\def\natexlab#1{#1}\fi
\expandafter\ifx\csname bibnamefont\endcsname\relax
  \def\bibnamefont#1{#1}\fi
\expandafter\ifx\csname bibfnamefont\endcsname\relax
  \def\bibfnamefont#1{#1}\fi
\expandafter\ifx\csname citenamefont\endcsname\relax
  \def\citenamefont#1{#1}\fi
\expandafter\ifx\csname url\endcsname\relax
  \def\url#1{\texttt{#1}}\fi
\expandafter\ifx\csname urlprefix\endcsname\relax\def\urlprefix{URL }\fi
\providecommand{\bibinfo}[2]{#2}
\providecommand{\eprint}[2][]{\url{#2}}

\bibitem[{\citenamefont{Klaers et~al.}(2010)\citenamefont{Klaers, Schmitt,
  Vewinger, and Weitz}}]{Klaers2010a}
\bibinfo{author}{\bibfnamefont{J.}~\bibnamefont{Klaers}},
  \bibinfo{author}{\bibfnamefont{J.}~\bibnamefont{Schmitt}},
  \bibinfo{author}{\bibfnamefont{F.}~\bibnamefont{Vewinger}}, \bibnamefont{and}
  \bibinfo{author}{\bibfnamefont{M.}~\bibnamefont{Weitz}},
  \bibinfo{journal}{Nature} \textbf{\bibinfo{volume}{468}},
  \bibinfo{pages}{545} (\bibinfo{year}{2010}).

\bibitem[{\citenamefont{Zou et~al.}(2004)\citenamefont{Zou, Janel, and
  Schatz}}]{Zou2004a}
\bibinfo{author}{\bibfnamefont{S.}~\bibnamefont{Zou}},
  \bibinfo{author}{\bibfnamefont{N.}~\bibnamefont{Janel}}, \bibnamefont{and}
  \bibinfo{author}{\bibfnamefont{G.~C.} \bibnamefont{Schatz}},
  \bibinfo{journal}{The Journal of Chemical Physics}
  \textbf{\bibinfo{volume}{120}}, \bibinfo{pages}{10871}
  (\bibinfo{year}{2004}).

\bibitem[{\citenamefont{Garcia~de Abajo}(2007)}]{GarciadeAbajo2007}
\bibinfo{author}{\bibfnamefont{F.~J.} \bibnamefont{Garcia~de Abajo}},
  \bibinfo{journal}{Rev. Mod. Phys.} \textbf{\bibinfo{volume}{79}},
  \bibinfo{pages}{1267} (\bibinfo{year}{2007}).

\bibitem[{\citenamefont{Augui{\'e} and Barnes}(2008)}]{Auguie2008}
\bibinfo{author}{\bibfnamefont{B.}~\bibnamefont{Augui{\'e}}} \bibnamefont{and}
  \bibinfo{author}{\bibfnamefont{W.~L.} \bibnamefont{Barnes}},
  \bibinfo{journal}{Phys. Rev. Lett.} \textbf{\bibinfo{volume}{101}},
  \bibinfo{pages}{143902} (\bibinfo{year}{2008}).

\bibitem[{\citenamefont{Bloch et~al.}(2008)\citenamefont{Bloch, Dalibard, and
  Zwerger}}]{Bloch2008}
\bibinfo{author}{\bibfnamefont{I.}~\bibnamefont{Bloch}},
  \bibinfo{author}{\bibfnamefont{J.}~\bibnamefont{Dalibard}}, \bibnamefont{and}
  \bibinfo{author}{\bibfnamefont{W.}~\bibnamefont{Zwerger}},
  \bibinfo{journal}{Rev. Mod. Phys.} \textbf{\bibinfo{volume}{80}},
  \bibinfo{pages}{885} (\bibinfo{year}{2008}).

\bibitem[{\citenamefont{Windpassinger and
  Sengstock}(2013)}]{windpassinger_engineering_2013}
\bibinfo{author}{\bibfnamefont{P.}~\bibnamefont{Windpassinger}}
  \bibnamefont{and}
  \bibinfo{author}{\bibfnamefont{K.}~\bibnamefont{Sengstock}},
  \bibinfo{journal}{Reports on Progress in Physics}
  \textbf{\bibinfo{volume}{76}}, \bibinfo{pages}{086401}
  (\bibinfo{year}{2013}).

\bibitem[{\citenamefont{T{\"{o}}rm{\"{a}} and
  Barnes}(2014)}]{torma_strong_2014}
\bibinfo{author}{\bibfnamefont{P.}~\bibnamefont{T{\"{o}}rm{\"{a}}}}
  \bibnamefont{and} \bibinfo{author}{\bibfnamefont{W.~L.}
  \bibnamefont{Barnes}}, \bibinfo{journal}{invited review to Rep.\ Prog.\
  Phys.\, arXiv:1405.1661}  (\bibinfo{year}{2014}).

\bibitem[{\citenamefont{Jaksch et~al.}(1998)\citenamefont{Jaksch, Bruder,
  Cirac, Gardiner, and Zoller}}]{jaksch_cold_1998}
\bibinfo{author}{\bibfnamefont{D.}~\bibnamefont{Jaksch}},
  \bibinfo{author}{\bibfnamefont{C.}~\bibnamefont{Bruder}},
  \bibinfo{author}{\bibfnamefont{J.~I.} \bibnamefont{Cirac}},
  \bibinfo{author}{\bibfnamefont{C.~W.} \bibnamefont{Gardiner}},
  \bibnamefont{and} \bibinfo{author}{\bibfnamefont{P.}~\bibnamefont{Zoller}},
  \bibinfo{journal}{Phys. Rev. Lett.} \textbf{\bibinfo{volume}{81}},
  \bibinfo{pages}{3108} (\bibinfo{year}{1998}).

\bibitem[{\citenamefont{Greiner et~al.}(2002)\citenamefont{Greiner, Mandel,
  Esslinger, H{\"{a}}nsch, and Bloch}}]{greiner_quantum_2002}
\bibinfo{author}{\bibfnamefont{M.}~\bibnamefont{Greiner}},
  \bibinfo{author}{\bibfnamefont{O.}~\bibnamefont{Mandel}},
  \bibinfo{author}{\bibfnamefont{T.}~\bibnamefont{Esslinger}},
  \bibinfo{author}{\bibfnamefont{T.~W.} \bibnamefont{H{\"{a}}nsch}},
  \bibnamefont{and} \bibinfo{author}{\bibfnamefont{I.}~\bibnamefont{Bloch}},
  \bibinfo{journal}{Nature} \textbf{\bibinfo{volume}{415}}, \bibinfo{pages}{39}
  (\bibinfo{year}{2002}), ISSN \bibinfo{issn}{0028-0836}.

\bibitem[{\citenamefont{Kasprzak et~al.}(2006)\citenamefont{Kasprzak, Richard,
  Kundermann, Baas, Jeambrun, Keeling, Marchetti, Szyma\'{n}ska, Andr\'{e},
  Staehli et~al.}}]{kasprzak_boseeinstein_2006}
\bibinfo{author}{\bibfnamefont{J.}~\bibnamefont{Kasprzak}},
  \bibinfo{author}{\bibfnamefont{M.}~\bibnamefont{Richard}},
  \bibinfo{author}{\bibfnamefont{S.}~\bibnamefont{Kundermann}},
  \bibinfo{author}{\bibfnamefont{A.}~\bibnamefont{Baas}},
  \bibinfo{author}{\bibfnamefont{P.}~\bibnamefont{Jeambrun}},
  \bibinfo{author}{\bibfnamefont{J.~M.~J.} \bibnamefont{Keeling}},
  \bibinfo{author}{\bibfnamefont{F.~M.} \bibnamefont{Marchetti}},
  \bibinfo{author}{\bibfnamefont{M.~H.} \bibnamefont{Szyma\'{n}ska}},
  \bibinfo{author}{\bibfnamefont{R.}~\bibnamefont{Andr\'{e}}},
  \bibinfo{author}{\bibfnamefont{J.~L.} \bibnamefont{Staehli}},
  \bibnamefont{et~al.}, \bibinfo{journal}{Nature}
  \textbf{\bibinfo{volume}{443}}, \bibinfo{pages}{409} (\bibinfo{year}{2006}).

\bibitem[{\citenamefont{Balili et~al.}(2007)\citenamefont{Balili, Hartwell,
  Snoke, Pfeiffer, and West}}]{balili_bose-einstein_2007}
\bibinfo{author}{\bibfnamefont{R.}~\bibnamefont{Balili}},
  \bibinfo{author}{\bibfnamefont{V.}~\bibnamefont{Hartwell}},
  \bibinfo{author}{\bibfnamefont{D.}~\bibnamefont{Snoke}},
  \bibinfo{author}{\bibfnamefont{L.}~\bibnamefont{Pfeiffer}}, \bibnamefont{and}
  \bibinfo{author}{\bibfnamefont{K.}~\bibnamefont{West}},
  \bibinfo{journal}{Science} \textbf{\bibinfo{volume}{316}},
  \bibinfo{pages}{1007} (\bibinfo{year}{2007}).

\bibitem[{\citenamefont{Plumhof et~al.}(2013)\citenamefont{Plumhof,
  St{\"{o}}ferle, Mai, Scherf, and Mahrt}}]{plumhof_room-temperature_2013}
\bibinfo{author}{\bibfnamefont{J.~D.} \bibnamefont{Plumhof}},
  \bibinfo{author}{\bibfnamefont{T.}~\bibnamefont{St{\"{o}}ferle}},
  \bibinfo{author}{\bibfnamefont{L.}~\bibnamefont{Mai}},
  \bibinfo{author}{\bibfnamefont{U.}~\bibnamefont{Scherf}}, \bibnamefont{and}
  \bibinfo{author}{\bibfnamefont{R.~F.} \bibnamefont{Mahrt}},
  \bibinfo{journal}{Nature Materials}  (\bibinfo{year}{2013}).

\bibitem[{\citenamefont{Yokoyama}(1992)}]{yokoyama_physics_1992}
\bibinfo{author}{\bibfnamefont{H.}~\bibnamefont{Yokoyama}},
  \bibinfo{journal}{Science} \textbf{\bibinfo{volume}{256}},
  \bibinfo{pages}{66} (\bibinfo{year}{1992}), \bibinfo{note}{{PMID:} 17802593}.

\bibitem[{\citenamefont{Rice and Carmichael}(1994)}]{rice_photon_1994}
\bibinfo{author}{\bibfnamefont{P.~R.} \bibnamefont{Rice}} \bibnamefont{and}
  \bibinfo{author}{\bibfnamefont{H.~J.} \bibnamefont{Carmichael}},
  \bibinfo{journal}{Phys. Rev. A} \textbf{\bibinfo{volume}{50}},
  \bibinfo{pages}{4318} (\bibinfo{year}{1994}).

\bibitem[{\citenamefont{Carusotto and Ciuti}(2013)}]{Carusotto2013}
\bibinfo{author}{\bibfnamefont{I.}~\bibnamefont{Carusotto}} \bibnamefont{and}
  \bibinfo{author}{\bibfnamefont{C.}~\bibnamefont{Ciuti}},
  \bibinfo{journal}{Rev. Mod. Phys.} \textbf{\bibinfo{volume}{85}},
  \bibinfo{pages}{299} (\bibinfo{year}{2013}).

\bibitem[{\citenamefont{Oulton et~al.}(2009)\citenamefont{Oulton, Sorger,
  Zentgraf, Ma, Gladden, Dai, Bartal, and Zhang}}]{oulton_plasmon_2009}
\bibinfo{author}{\bibfnamefont{R.~F.} \bibnamefont{Oulton}},
  \bibinfo{author}{\bibfnamefont{V.~J.} \bibnamefont{Sorger}},
  \bibinfo{author}{\bibfnamefont{T.}~\bibnamefont{Zentgraf}},
  \bibinfo{author}{\bibfnamefont{R.-M.} \bibnamefont{Ma}},
  \bibinfo{author}{\bibfnamefont{C.}~\bibnamefont{Gladden}},
  \bibinfo{author}{\bibfnamefont{L.}~\bibnamefont{Dai}},
  \bibinfo{author}{\bibfnamefont{G.}~\bibnamefont{Bartal}}, \bibnamefont{and}
  \bibinfo{author}{\bibfnamefont{X.}~\bibnamefont{Zhang}},
  \bibinfo{journal}{Nature} \textbf{\bibinfo{volume}{461}},
  \bibinfo{pages}{629} (\bibinfo{year}{2009}).

\bibitem[{\citenamefont{Zhou et~al.}(2013)\citenamefont{Zhou, Dridi, Suh, Kim,
  Co, Wasielewski, Schatz, and Odom}}]{zhou_lasing_2013}
\bibinfo{author}{\bibfnamefont{W.}~\bibnamefont{Zhou}},
  \bibinfo{author}{\bibfnamefont{M.}~\bibnamefont{Dridi}},
  \bibinfo{author}{\bibfnamefont{J.~Y.} \bibnamefont{Suh}},
  \bibinfo{author}{\bibfnamefont{C.~H.} \bibnamefont{Kim}},
  \bibinfo{author}{\bibfnamefont{D.~T.} \bibnamefont{Co}},
  \bibinfo{author}{\bibfnamefont{M.~R.} \bibnamefont{Wasielewski}},
  \bibinfo{author}{\bibfnamefont{G.~C.} \bibnamefont{Schatz}},
  \bibnamefont{and} \bibinfo{author}{\bibfnamefont{T.~W.} \bibnamefont{Odom}},
  \bibinfo{journal}{Nature Nanotechnology} \textbf{\bibinfo{volume}{8}},
  \bibinfo{pages}{506} (\bibinfo{year}{2013}).

\bibitem[{\citenamefont{van Beijnum et~al.}(2013)\citenamefont{van Beijnum, van
  Veldhoven, Geluk, de~Dood, 't~Hooft, and van Exter}}]{vanBeijum2013}
\bibinfo{author}{\bibfnamefont{F.}~\bibnamefont{van Beijnum}},
  \bibinfo{author}{\bibfnamefont{P.~J.} \bibnamefont{van Veldhoven}},
  \bibinfo{author}{\bibfnamefont{E.~J.} \bibnamefont{Geluk}},
  \bibinfo{author}{\bibfnamefont{M.~J.~A.} \bibnamefont{de~Dood}},
  \bibinfo{author}{\bibfnamefont{G.~W.} \bibnamefont{'t~Hooft}},
  \bibnamefont{and} \bibinfo{author}{\bibfnamefont{M.~P.} \bibnamefont{van
  Exter}}, \bibinfo{journal}{Phys. Rev. Lett.} \textbf{\bibinfo{volume}{110}},
  \bibinfo{pages}{206802} (\bibinfo{year}{2013}).

\bibitem[{\citenamefont{Rodriguez and {G\'{o}mez Rivas}}(2013)}]{Rodriguez2013}
\bibinfo{author}{\bibfnamefont{S.}~\bibnamefont{Rodriguez}} \bibnamefont{and}
  \bibinfo{author}{\bibfnamefont{J.}~\bibnamefont{{G\'{o}mez Rivas}}},
  \bibinfo{journal}{Optics Express} \textbf{\bibinfo{volume}{21}},
  \bibinfo{pages}{27411} (\bibinfo{year}{2013}).

\bibitem[{\citenamefont{V{\"{a}}kev{\"{a}}inen
  et~al.}(2013)\citenamefont{V{\"{a}}kev{\"{a}}inen, Moerland, Rekola,
  Eskelinen, Martikainen, Kim, and
  T{\"{o}}rm{\"{a}}}}]{vakevainen_plasmonic_2013}
\bibinfo{author}{\bibfnamefont{A.~I.} \bibnamefont{V{\"{a}}kev{\"{a}}inen}},
  \bibinfo{author}{\bibfnamefont{R.~J.} \bibnamefont{Moerland}},
  \bibinfo{author}{\bibfnamefont{H.~T.} \bibnamefont{Rekola}},
  \bibinfo{author}{\bibfnamefont{A.-P.} \bibnamefont{Eskelinen}},
  \bibinfo{author}{\bibfnamefont{J.-P.} \bibnamefont{Martikainen}},
  \bibinfo{author}{\bibfnamefont{D.-H.} \bibnamefont{Kim}}, \bibnamefont{and}
  \bibinfo{author}{\bibfnamefont{P.}~\bibnamefont{T{\"{o}}rm{\"{a}}}},
  \bibinfo{journal}{Nano Letters} \textbf{\bibinfo{volume}{14}},
  \bibinfo{pages}{1721–1727} (\bibinfo{year}{2013}).

\bibitem[{\citenamefont{Shi et~al.}(2014)\citenamefont{Shi, Hakala, Rekola,
  Martikainen, Moerland, and T\"{o}rm\"{a}}}]{shi_spatial_2014}
\bibinfo{author}{\bibfnamefont{L.}~\bibnamefont{Shi}},
  \bibinfo{author}{\bibfnamefont{T.}~\bibnamefont{Hakala}},
  \bibinfo{author}{\bibfnamefont{H.}~\bibnamefont{Rekola}},
  \bibinfo{author}{\bibfnamefont{J.-P.} \bibnamefont{Martikainen}},
  \bibinfo{author}{\bibfnamefont{R.}~\bibnamefont{Moerland}}, \bibnamefont{and}
  \bibinfo{author}{\bibfnamefont{P.}~\bibnamefont{T\"{o}rm\"{a}}},
  \bibinfo{journal}{Phys. Rev. Lett.} \textbf{\bibinfo{volume}{112}},
  \bibinfo{pages}{153002} (\bibinfo{year}{2014}).

\bibitem[{\citenamefont{Fischer and Weill}(2012)}]{fischer_when_2012}
\bibinfo{author}{\bibfnamefont{B.}~\bibnamefont{Fischer}} \bibnamefont{and}
  \bibinfo{author}{\bibfnamefont{R.}~\bibnamefont{Weill}},
  \bibinfo{journal}{Optics Express} \textbf{\bibinfo{volume}{20}},
  \bibinfo{pages}{26704} (\bibinfo{year}{2012}).

\bibitem[{\citenamefont{Kirton and Keeling}(2013)}]{kirton_nonequilibrium_2013}
\bibinfo{author}{\bibfnamefont{P.}~\bibnamefont{Kirton}} \bibnamefont{and}
  \bibinfo{author}{\bibfnamefont{J.}~\bibnamefont{Keeling}},
  \bibinfo{journal}{Phys. Rev. Lett.} \textbf{\bibinfo{volume}{111}},
  \bibinfo{pages}{100404} (\bibinfo{year}{2013}).

\bibitem[{\citenamefont{Yamamoto et~al.}(1991)\citenamefont{Yamamoto, Machida,
  and Bj{\"{o}}rk}}]{yamamoto_microcavity_1991}
\bibinfo{author}{\bibfnamefont{Y.}~\bibnamefont{Yamamoto}},
  \bibinfo{author}{\bibfnamefont{S.}~\bibnamefont{Machida}}, \bibnamefont{and}
  \bibinfo{author}{\bibfnamefont{G.}~\bibnamefont{Bj{\"{o}}rk}},
  \bibinfo{journal}{Phys. Rev. A} \textbf{\bibinfo{volume}{44}},
  \bibinfo{pages}{657} (\bibinfo{year}{1991}).

\bibitem[{\citenamefont{Moskovits}(1985)}]{moskovits_surface-enhanced_1985}
\bibinfo{author}{\bibfnamefont{M.}~\bibnamefont{Moskovits}},
  \bibinfo{journal}{Rev. Mod. Phys.} \textbf{\bibinfo{volume}{57}},
  \bibinfo{pages}{783} (\bibinfo{year}{1985}).

\bibitem[{\citenamefont{Klaers et~al.}(2011{\natexlab{a}})\citenamefont{Klaers,
  Schmitt, Damm, Vewinger, and Weitz}}]{klaers_boseeinstein_2011}
\bibinfo{author}{\bibfnamefont{J.}~\bibnamefont{Klaers}},
  \bibinfo{author}{\bibfnamefont{J.}~\bibnamefont{Schmitt}},
  \bibinfo{author}{\bibfnamefont{T.}~\bibnamefont{Damm}},
  \bibinfo{author}{\bibfnamefont{F.}~\bibnamefont{Vewinger}}, \bibnamefont{and}
  \bibinfo{author}{\bibfnamefont{M.}~\bibnamefont{Weitz}},
  \bibinfo{journal}{Applied Physics B} \textbf{\bibinfo{volume}{105}},
  \bibinfo{pages}{17} (\bibinfo{year}{2011}{\natexlab{a}}).

\bibitem[{\citenamefont{Rodriguez et~al.}(2011)\citenamefont{Rodriguez, Abass,
  Maes, Janssen, Vecchi, and G\'{o}mez~Rivas}}]{rodriguez_coupling_2011}
\bibinfo{author}{\bibfnamefont{S.~R.~K.} \bibnamefont{Rodriguez}},
  \bibinfo{author}{\bibfnamefont{A.}~\bibnamefont{Abass}},
  \bibinfo{author}{\bibfnamefont{B.}~\bibnamefont{Maes}},
  \bibinfo{author}{\bibfnamefont{O.~T.~A.} \bibnamefont{Janssen}},
  \bibinfo{author}{\bibfnamefont{G.}~\bibnamefont{Vecchi}}, \bibnamefont{and}
  \bibinfo{author}{\bibfnamefont{J.}~\bibnamefont{G\'{o}mez~Rivas}},
  \bibinfo{journal}{Phys. Rev. X} \textbf{\bibinfo{volume}{1}},
  \bibinfo{pages}{021019} (\bibinfo{year}{2011}).

\bibitem[{\citenamefont{Klaers et~al.}(2011{\natexlab{b}})\citenamefont{Klaers,
  Schmitt, Damm, Vewinger, and Weitz}}]{Klaers2011a}
\bibinfo{author}{\bibfnamefont{J.}~\bibnamefont{Klaers}},
  \bibinfo{author}{\bibfnamefont{J.}~\bibnamefont{Schmitt}},
  \bibinfo{author}{\bibfnamefont{T.}~\bibnamefont{Damm}},
  \bibinfo{author}{\bibfnamefont{F.}~\bibnamefont{Vewinger}}, \bibnamefont{and}
  \bibinfo{author}{\bibfnamefont{M.}~\bibnamefont{Weitz}},
  \bibinfo{journal}{Applied Physics B} \textbf{\bibinfo{volume}{105}},
  \bibinfo{pages}{17} (\bibinfo{year}{2011}{\natexlab{b}}).

\bibitem[{\citenamefont{Pethick and Smith}(2008)}]{pethick_bose-einstein_2008}
\bibinfo{author}{\bibfnamefont{C.}~\bibnamefont{Pethick}} \bibnamefont{and}
  \bibinfo{author}{\bibfnamefont{H.}~\bibnamefont{Smith}},
  \emph{\bibinfo{title}{Bose-Einstein condensation in dilute gases}}
  (\bibinfo{publisher}{Cambridge University Press},
  \bibinfo{address}{Cambridge; New York}, \bibinfo{year}{2008}).

\bibitem[{\citenamefont{Schmitt et~al.}(2014)\citenamefont{Schmitt, Damm, Dung,
  Vewinger, Klaers, and Weitz}}]{schmitt_observation_2014}
\bibinfo{author}{\bibfnamefont{J.}~\bibnamefont{Schmitt}},
  \bibinfo{author}{\bibfnamefont{T.}~\bibnamefont{Damm}},
  \bibinfo{author}{\bibfnamefont{D.}~\bibnamefont{Dung}},
  \bibinfo{author}{\bibfnamefont{F.}~\bibnamefont{Vewinger}},
  \bibinfo{author}{\bibfnamefont{J.}~\bibnamefont{Klaers}}, \bibnamefont{and}
  \bibinfo{author}{\bibfnamefont{M.}~\bibnamefont{Weitz}},
  \bibinfo{journal}{Phys. Rev. Lett.} \textbf{\bibinfo{volume}{112}},
  \bibinfo{pages}{030401} (\bibinfo{year}{2014}).

\bibitem[{\citenamefont{Altman et~al.}(2013)\citenamefont{Altman, Toner,
  Sieberer, Diehl, and Chen}}]{altman_two-dimensional_2013}
\bibinfo{author}{\bibfnamefont{E.}~\bibnamefont{Altman}},
  \bibinfo{author}{\bibfnamefont{J.}~\bibnamefont{Toner}},
  \bibinfo{author}{\bibfnamefont{L.~M.} \bibnamefont{Sieberer}},
  \bibinfo{author}{\bibfnamefont{S.}~\bibnamefont{Diehl}}, \bibnamefont{and}
  \bibinfo{author}{\bibfnamefont{L.}~\bibnamefont{Chen}},
  \bibinfo{type}{{arXiv} e-print} \bibinfo{number}{1311.0876}
  (\bibinfo{year}{2013}).

\bibitem[{\citenamefont{Mandel and Wolf}(1995)}]{mandel_optical_1995}
\bibinfo{author}{\bibfnamefont{L.}~\bibnamefont{Mandel}} \bibnamefont{and}
  \bibinfo{author}{\bibfnamefont{E.}~\bibnamefont{Wolf}},
  \emph{\bibinfo{title}{Optical Coherence and Quantum Optics}}
  (\bibinfo{publisher}{Cambridge University Press}, \bibinfo{address}{Cambridge
  ; New York}, \bibinfo{year}{1995}), \bibinfo{edition}{1st} ed., ISBN
  \bibinfo{isbn}{9780521417112}.

\bibitem[{\citenamefont{Grynberg et~al.}(2010)\citenamefont{Grynberg, Aspect,
  Fabre, and Cohen-Tannoudji}}]{grynberg_introduction_2010}
\bibinfo{author}{\bibfnamefont{G.}~\bibnamefont{Grynberg}},
  \bibinfo{author}{\bibfnamefont{A.}~\bibnamefont{Aspect}},
  \bibinfo{author}{\bibfnamefont{C.}~\bibnamefont{Fabre}}, \bibnamefont{and}
  \bibinfo{author}{\bibfnamefont{C.}~\bibnamefont{Cohen-Tannoudji}},
  \emph{\bibinfo{title}{Introduction to Quantum Optics: From the Semi-classical
  Approach to Quantized Light}} (\bibinfo{publisher}{Cambridge University
  Press}, \bibinfo{address}{Cambridge, {UK} ; New York}, \bibinfo{year}{2010}).

\end{thebibliography}

\end{document}